\documentclass[utf8]{FrontiersinHarvard} % for articles in journals using the Harvard Referencing Style (Author-Date), for Frontiers Reference Styles by Journal: https://zendesk.frontiersin.org/hc/en-us/articles/360017860337-Frontiers-Reference-Styles-by-Journal
\usepackage{url, hyperref, lineno} % microtype, subcaption}
\usepackage[onehalfspacing]{setspace}
\usepackage{doi}
\usepackage{multirow, tabularx}

\graphicspath{ {figures/} }

\def\keyFont{\fontsize{8}{11}\helveticabold }
\def\firstAuthorLast{Palmroos {et~al.}} %use et al only if is more than 1 author
\def\Authors{Christian Palmroos\,$^{1,*}$,
Jan Gieseler\,$^{1}$,
Nina Dresing\,$^{1}$,
Diana E. Morosan\,$^{2}$,
Eleanna Asvestari\,$^{2}$,
Aleksi Yli-Laurila\,$^{1}$,
Daniel J. Price\,$^{2}$,
Saku Valkila\,$^{1}$,
Rami Vainio\,$^{1}$
}

\def\Address{
$^{1}$Space Research Laboratory, Department of Physics and Astronomy, University of Turku, Turku, Finland\\
$^{2}$Department of Physics, University of Helsinki, P.O. Box 64, FI-00014 Helsinki, Finland
}
\def\corrAuthor{Christian Palmroos, University of Turku, 20500 Turku, Finland}

\def\corrEmail{christian.o.palmroos@utu.fi}

\begin{document}
\onecolumn
\firstpage{1}

\title[SEP Analysis]{Solar Energetic Particle Time Series Analysis with Python}

\author[\firstAuthorLast ]{\Authors} %This field will be automatically populated
\address{} %This field will be automatically populated
\correspondance{} %This field will be automatically populated

\extraAuth{}% If there are more than 1 corresponding author, comment this line and uncomment the next one.
%\extraAuth{corresponding Author2 \\ Laboratory X2, Institute X2, Department X2, Organization X2, Street X2, City X2 , State XX2 (only USA, Canada and Australia), Zip Code2, X2 Country X2, email2@uni2.edu}

\maketitle

\begin{abstract}

Solar Energetic Particles (SEPs) are charged particles accelerated within the solar atmosphere or the interplanetary space by explosive phenomena such as solar flares or Coronal Mass Ejections (CMEs). Once injected into the interplanetary space, they can propagate towards Earth, causing space weather related phenomena.
For their analysis, interplanetary in-situ measurements of charged particles are key. The recently expanded spacecraft fleet in the heliosphere not only provides much-needed additional vantage points, but also increases the variety of missions and instruments for which data loading and processing tools are needed. 
This manuscript introduces a series of Python functions that will enable the scientific community to download, load, and visualize charged particle measurements of the current space missions that are especially relevant to particle research as time series or dynamic spectra. In addition, further analytical functionality is provided that allows the determination of SEP onset times as well as their inferred injection times. The full workflow, which is intended to be run within Jupyter Notebooks and can also be approachable for Python laymen, will be presented with scientific examples. All functions are written in Python, with the source code publicly available at GitHub under a permissive license. Where appropriate, available Python libraries are used, and their application is described.

%%% Leave the Abstract empty if your article does not require one, please see the Summary Table for full details.
%\section{}
% For full guidelines regarding your manuscript please refer to \href{http://www.frontiersin.org/about/AuthorGuidelines}{Author Guidelines}.

%As a primary goal, the abstract should render the general significance and conceptual advance of the work clearly accessible to a broad readership. References should not be cited in the abstract. Leave the Abstract empty if your article does not require one, please see \href{http://www.frontiersin.org/about/AuthorGuidelines#SummaryTable}{Summary Table} for details according to article type. 

\tiny
 \keyFont{ \section{Keywords:} Python, Software Package, Solar Energetic Particle (SEP), Coronal Mass Ejection (CME), Spacecraft, Heliosphere, Data, Onset Time} %All article types: you may provide up to 8 keywords; at least 5 are mandatory.
\end{abstract}

\section{Introduction}
% Short general intro to SEPs. Motivation for tools developed here.
Solar Energetic Particle (SEP) events, as observed at a spacecraft, are determined by a combination of physical processes such as their acceleration mechanism, the particle injection into the interplanetary magnetic field, and their transport through this medium \citep[see, e.g.,][and reference therein for a review]{Desai2016, Reames2021}. All of these processes can vary, and they are not yet fully understood. Using only measurements of a single spacecraft, it can be very difficult to disentangle these processes, making multi-spacecraft observations indispensable. A careful and comprehensive analysis is therefore needed. It should focus not only on the energetic particle observations but also on in-situ measurements of the magnetic field and solar wind plasma, as well as on remote-sensing observations of the solar counterpart of the event, such as a solar flare or a Coronal Mass Ejection (CME), both being sites of potential particle acceleration. 

A powerful method to study SEP events is to use multi-spacecraft observations of the same event that are situated at various locations with respect to the source region. This can allow, for example, to disentangle source properties from transport effects. The tools presented in this manuscript, which enable such multi-spacecraft analyses, have been developed within the \emph{Solar energetic particle analysis platform for the inner heliosphere} (\href{https://serpentine-h2020.eu/}{SERPENTINE}), funded by the European Union's Horizon 2020 framework programme (see \citealp{Gieseler2022} for an introduction).
They focus mainly on the study of an SEP event, which requires a careful characterization of its various features. These include the shape of the energetic particle time profiles at various energies and for different particle species, the maximum observed energies, as well as peak intensities, differences in directional measurements that allow determining anisotropies, and the onset time of the event. The latter is defined by the first arriving particles at the spacecraft, and it is one of the most important parameters for linking the energetic particle measurements with remote-sensing observations of the solar phenomena associated to the SEP event. As the kinetic energy, and hence the the speed, of the detected particles is usually known, one can infer their injection time at the Sun when assuming a certain path length they had to travel to reach the spacecraft. This allows then the comparison with the times of various solar phenomena to figure out their importance for the SEP event. Considering that higher-energy particles travel faster than lower-energy ones, the onset of a SEP event often shows a velocity dispersion, i.e., higher-energy particles arriving earlier than those with lower energies. An analysis of this velocity dispersion, together with the assumption that all particles were injected at the same time and travelled the same path length, allows inferring not only a common injection time but also the path length itself.

The tools presented here constitute a comprehensive toolkit that enables the user to perform the aforementioned energetic particle characterization using multiple spacecraft, such as Solar Orbiter \citep{Mueller2020}, Parker Solar Probe \citep{Fox2016}, Solar Terrestrial Relations Observatory \citep[STEREO, ][]{Kaiser2005,Kaiser2008}, SOlar and Heliospheric Observatory \citep[SOHO,][]{Domingo95}, and Wind \citep{Harten1995}. The basis of the toolkit lies in the functionality that allows the user to automatically download data files from online repositories and to load them into powerful Pandas DataFrame objects \citep{Mckinney2010} in Python. All the data loaders (cf. Sect.~\ref{sect:data_loaders}) are made available through a Python package that can be installed with a single command. In addition, a Jupyter Notebook provides visualization examples and additional functionality to explore the content of the loaded data files, for example various energy ranges, particle species and further metadata. 
The rest of the tools presented in this paper use the data loaders in the background and provide more advanced analysis methods, such as the onset determination tool (Sect.~\ref{sect:onsets}) or the visual time shift analysis tool (Sect.~\ref{sect:tsa}). Finally, we also provide a tool that allows a first comparison of another phenomenon of solar activity, namely radio bursts, with the energetic particle observations, the spectrogram tool (Sect.~\ref{sect:spectra}). It can be used to plot dynamic spectrograms of energetic particles as well as radio observations.

\section{Method} % / Analysis and Visualization tools}
All the tools described in this paper are open-source Python software released under the BSD-3-Clause license. In general, they are all provided as self-explanatory Jupyter Notebooks (\citealp{Kluyver2016}; see \href{https://jupyter.org}{jupyter.org} for more details) that give detailed instructions and examples (see Figs.~\ref{fig:nb_dataloader} and \ref{fig:nb_onset_input} for examples). Based on these Notebooks, the user should be able to modify the code according to their needs. 
The actual software code of the tools is delivered as a Python package called \texttt{SEPpy} \citep{seppy2022}. It is listed on \href{https://pypi.org/project/solarmach/}{PyPI} so that it can be installed using the widely-used \texttt{pip} command. By delivering the package in this manner, all other required packages are automatically installed.
In the Notebooks itself, all necessary functions are imported from this package. Thus, the user only needs to modify some parameters to their needs, such as selecting the date or instrument of interest. 
Depending on this selection, all required data files are dynamically loaded from online sources or local files using the data loaders presented in Sect.~\ref{sect:data_loaders}.
The Jupyter Notebooks are distributed through the GitHub repository \href{https://github.com/serpentine-h2020/serpentine/}{serpentine-h2020/serpentine} \citep{SERPENTINE2022}, which is archived at Zenodo \citep{Zenodo}. The repository contains detailed installation instructions, and the tools will be continuously updated through it.

While running the tools as Jupyter Notebooks requires only a very basic level of Python knowledge, the user still needs to follow an installation procedure in order to use the tools. As an alternative, the SERPENTINE project provides a JupyterHub server that allows users to run the Notebooks described in this manuscript completely online, without any requirements except for a web browser and a GitHub account for authentication. More details about the JupyterHub server can be found on the \href{https://serpentine-h2020.eu/hub/}{SERPENTINE website} and will be presented in an upcoming publication.

\subsection{Data loaders} 
\label{sect:data_loaders}
The Data loaders Notebook is a collection of functions that drastically simplifies obtaining (i.e., automatically downloading and loading into Python structures) energetic charged particle data sets of in-situ measurements by the current heliospheric spacecraft fleet. This is especially valuable because the data products of the newest spacecraft like Solar Orbiter or Parker Solar Probe are only released in the binary Common Data Format (CDF) that requires advanced programming skills in order to read just the basic data structures. 
\begin{table}[ht!]
\caption{All instruments supported by the Energetic\\Particle Data Loaders Notebook.}
\label{tab:data_loaders}
\begin{tabularx}{0.5\textwidth}{ll}
\toprule
\textbf{Spacecraft}                    & \textbf{Instrument} \\ 
\midrule
\multirow[t]{2}{*}{Parker Solar Probe} & IS$\odot$IS/EPI-Hi HET \\
                                       & IS$\odot$IS/EPI-Lo PE  \\
\multirow[t]{5}{*}{SOHO}               & COSTEP/EPHIN     \\
                                       & ERNE-HED         \\
                                       & ERNE-LED         \\
\multirow[t]{3}{*}{Solar Orbiter}      & EPD/EPT          \\
                                       & EPD/HET          \\
                                       & EPD/STEP$^1$     \\
\multirow[t]{3}{*}{STEREO-A/B$^2$}     & HET              \\
                                       & LET              \\
                                       & SEPT             \\
Wind                                   & 3DP              \\ 
\toprule
\multicolumn{2}{l}{\small $^1$so far only until Oct 2021 due to a change in data product} \\
\multicolumn{2}{l}{\small $^2$STEREO-B only until Oct 2014 due to spacecraft contact loss} \\
\end{tabularx}
\end{table}
Table~\ref{tab:data_loaders} gives an overview of the currently supported instruments. For most of the missions, the data is downloaded as CDF data files from the Coordinated Data Analysis Web (\href{https://cdaweb.gsfc.nasa.gov}{CDAWeb}) service provided by NASA's Space Physics Data Facility at Goddard Space Flight Center. The data loading is in these cases utilized through version 4.0.5 \citep{Mumford2022} of the \texttt{sunpy} open-source software package \citep{Sunpy2020}. Some data sets are directly obtained from the web servers of the responsible institutes, primarily due to not being available through CDAWeb. This applies to Wind/3DP data, STEREO/SEPT data, and electron measurements of SOHO/EPHIN. The latter two data sets are obtained as ASCII files, and not in the CDF format. 
In general, level 2 data products or higher are used. This is calibrated and validated data that is ready for scientific use \citep[e.g.,][]{McComas2016,Rodriguez-Pacheco2020}. However, in some cases, when for example only level 1 data is available, lower-level data products are chosen.
All data files from the Energetic Particle Detector (EPD) instrument suite of Solar Orbiter are directly obtained as CDF files from ESA's official \href{http://soar.esac.esa.int/soar/}{Solar Orbiter Archive (SOAR)}, using its own Application Programming Interface (API). For this, the already established version 0.1.10 \citep{solo-epd-loader2022} of the \texttt{solo-epd-loader} Python package is used, which in addition to level 2 data also supports the so-called low-latency (LL) data. This data product provides the latest observations, which are very useful in terms of quick-look purposes. However, because the data set is not yet verified it  shall not be used for scientific publications. 

\begin{figure}[hp!]
\begin{center}
\includegraphics[height=0.94\textheight]{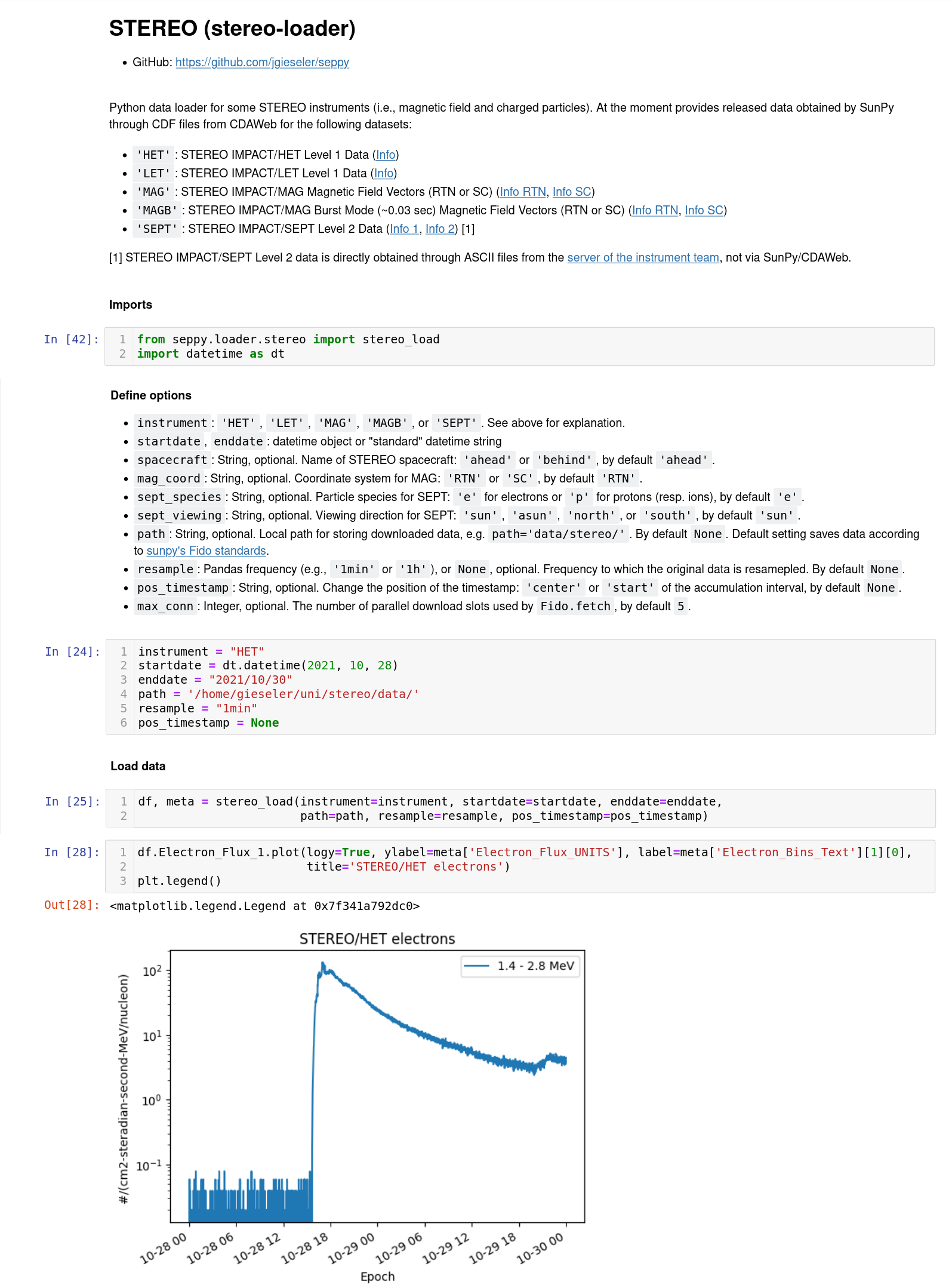}
\end{center}
\caption{Example workflow within the Data loaders Notebook showing how to load and visualize STEREO/HET observations.} \label{fig:nb_dataloader}
\end{figure}

For most data sets, the data files are loaded into Pandas DataFrames \citep{Mckinney2010} in Python using SunPy's \href{https://docs.sunpy.org/en/stable/guide/data_types/timeseries.html}{\texttt{TimeSeries}} functionality. In some cases this is not possible, namely when the data is multidimensional, that is, if multiple dependencies are explicitly defined in the corresponding CDF file. This can for example be the case if a measurement is not only related to time, but also to energy and viewing direction, and depends on the way the CDF file has been structured by the instrument team. In these cases, manual loading functions adopted from the function \texttt{cdf2df} of version 0.15.4 of the discontinued Python package \texttt{heliopy} \citep{Heliopy} are used. These functions exploit the Python package \texttt{cdflib} \citep{cdflib}.

Figure~\ref{fig:nb_dataloader} shows an example workflow within the Data loaders Notebook that illustrates how to load data from the STEREO mission. Next to a direct example, the Notebook also gives information on the other available options.
\begin{figure}[ht!]
\centering
\includegraphics[width=\textwidth]{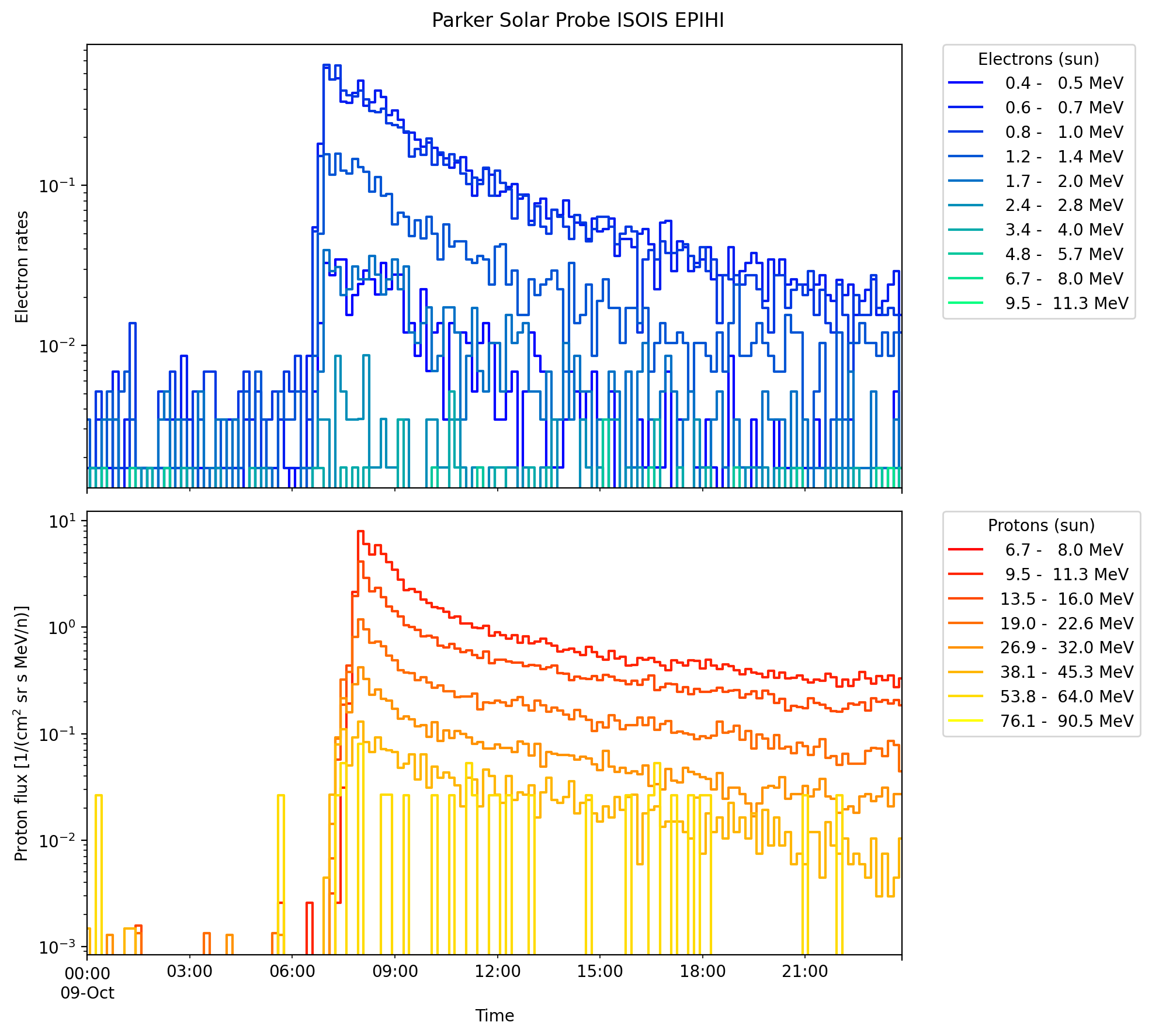}
  \caption{Example output of plotting Parker Solar Probe data in the SEP Data Loaders Notebook. Shown here are electron and proton measurements of the IS$\odot$IS/EPIHI instrument of Parker Solar Probe for the SEP event on 9 October 2021. Note that the electron measurements are count rates and not fluxes.
          }
     \label{fig:data_loader_psp}
\end{figure}
\begin{figure}[ht!]
\centering
\includegraphics[width=\textwidth]{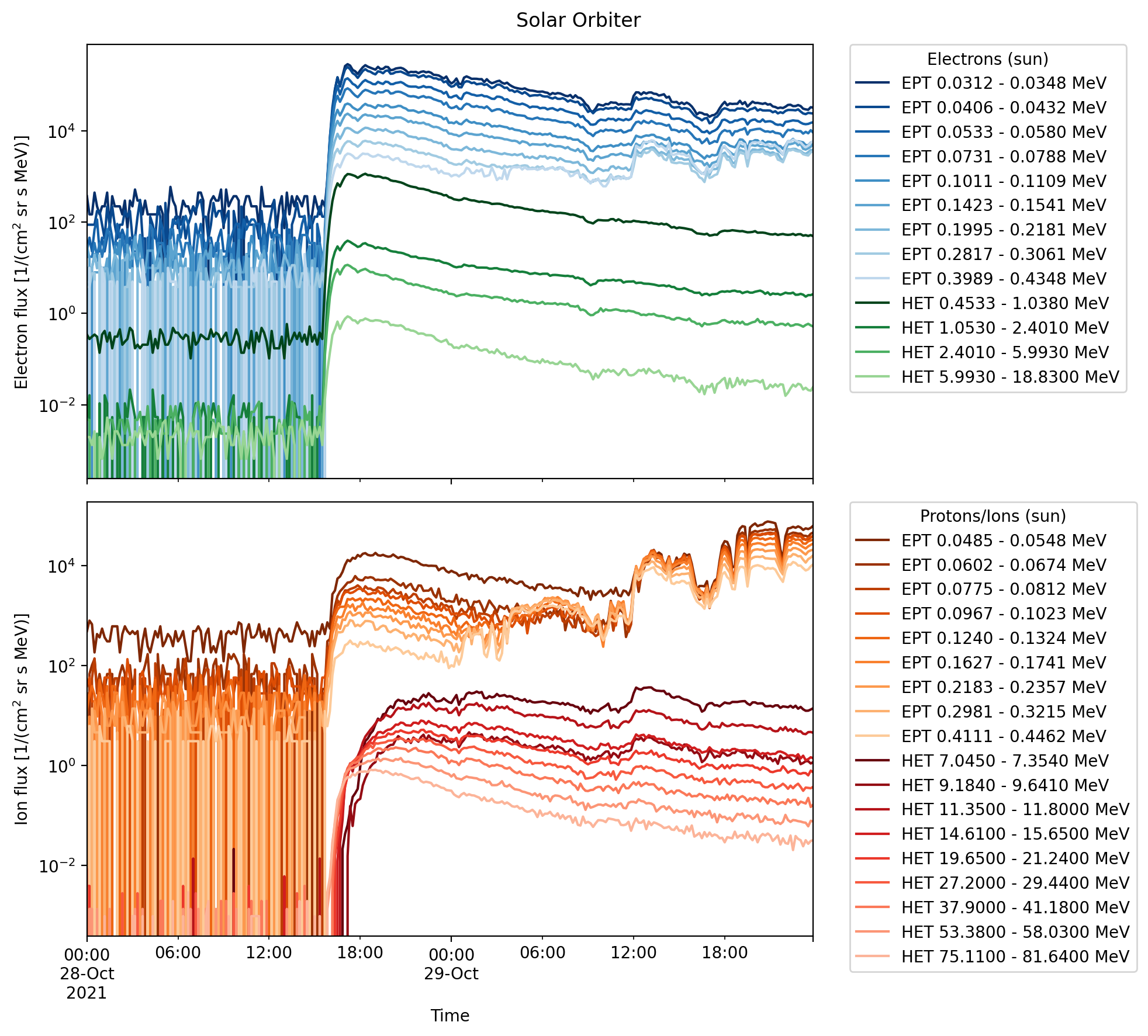}
  \caption{Example output of how to combine different instrument observations into a multi-panel plot in the SEP Data Loaders Notebook. Shown here are electron and ion measurements of the EPT and HET sensors of Solar Orbiter's EPD instrument suite for the SEP event on 29 October 2021. Note that the increases in ion fluxes observed by EPT around 18:00 are due to contamination caused by electrons.
          }
     \label{fig:data_loader_1}
\end{figure}
\begin{figure}[ht!]
\centering
\includegraphics[width=\textwidth]{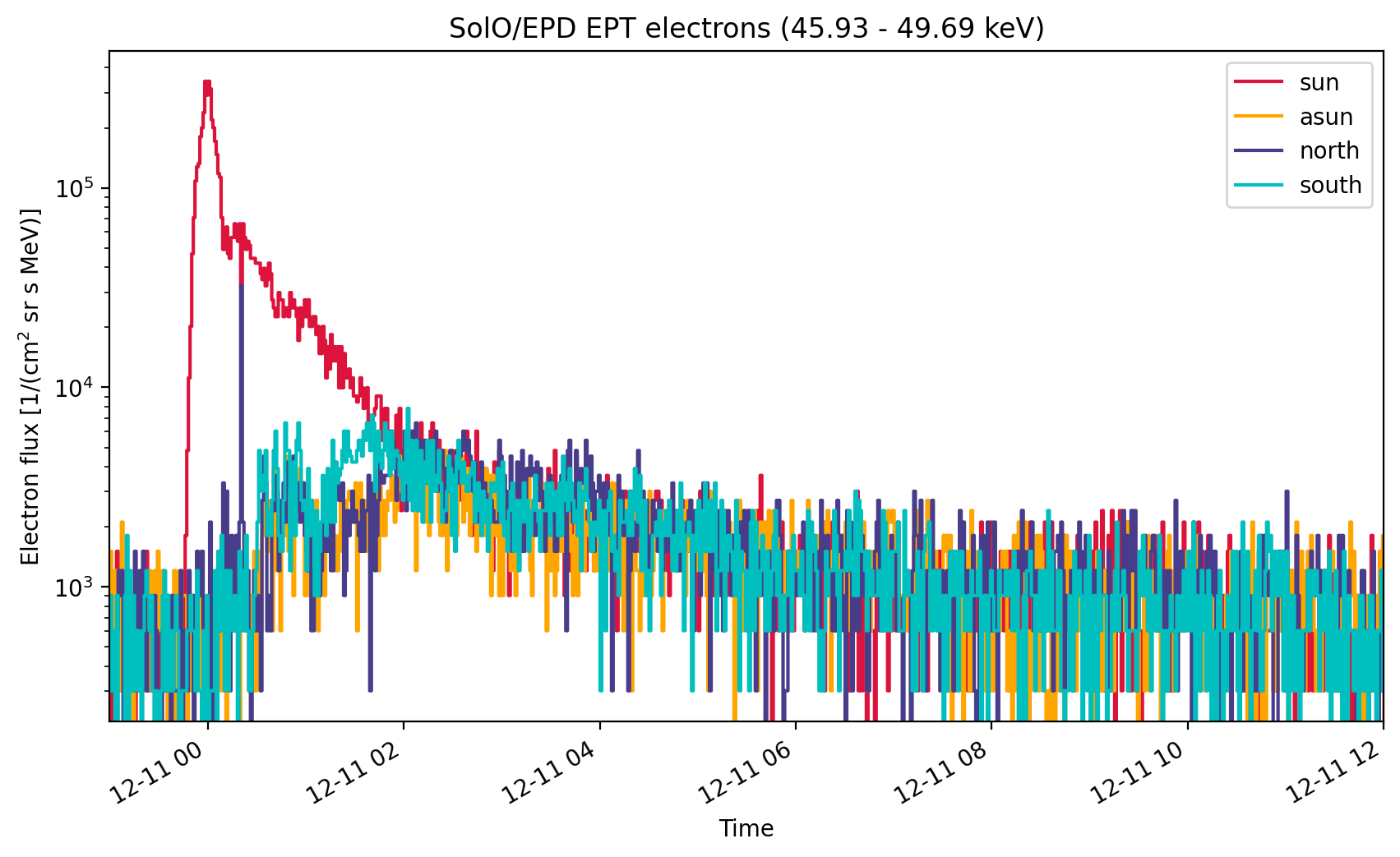}
  \caption{Example output of how to add multiple viewing directions into a single plot in the Data loaders Notebook. Shown here are the four viewing directions of the EPT sensor of Solar Orbiter's EPD instrument suite.}
     \label{fig:data_loader_2}
\end{figure}
In addition, the Notebook provides different examples on how to visualize the data with Python. For example, how to combine observations of multiple energy channels of different species (see Fig.~\ref{fig:data_loader_psp}) or 
different instruments into a multi-panel plot (Fig.~\ref{fig:data_loader_1}), or how to add multiple viewing directions into a single plot (Fig.~\ref{fig:data_loader_2}). Versed Python users can then easily build their own analyses based on this. With these functions, they can skip the inconvenient start of setting up corresponding routines, and can directly start their scientific analysis. 
This Jupyter Notebook shows only a selection of data loading examples for some of the data loaders, and will be continuously expanded with further missions and instruments. 
The data loading functionality of this Notebook is employed also in the following Notebooks presented in Sects.~\ref{sect:onsets} to \ref{sect:tsa}.

Next to the loading of charged particle observations, in SEPpy partial support is given for magnetic field and radio measurements. For these, the download and loading of the data is done in the same way as described above. At the moment, the data obtained by the MAG instruments onboard STEREO and Solar Orbiter as well as radio observations by the STEREO/SWAVES instrument are included. The radio data is especially important for the dynamic spectrum tool described in Sect.~\ref{sect:spectra}.

\subsection{SEP event onset determination}
\label{sect:onsets}
The onset time of an SEP event, that is, the initially measured intensity increase at a spacecraft, is a key information for the event analysis. For example, it is necessary for inferring the injection time and connecting remote-sensing observations to in-situ measurements. Furthermore, by knowing the onset times of particles of varying energies, one can get a better estimation of the path that the particles travelled. As the particles travel at different speeds corresponding to their different kinetic energies, they are expected to show different onset times and profiles for different energy channels if one assumes that all particles were injected at the same time at the Sun. This effect is known as velocity dispersion, and it is the basis for a Velocity Dispersion Analysis \citep[VDA, e.g.,][]{lintunen_vainio_vda}, in which one can derive the initial injection of the SEPs at the Sun. There are different methods to determine onset times, such as the widely used 3-sigma method, in which the event is defined by the time when a certain number of data points increase by 3 sigma above the predefined background, or by applying a fit to the rising phase of the event in order to extrapolate backwards \citep[e.g., ][]{Dresing2012}. The tool presented in this paper employs the Poisson-CUSUM method \citep{lucas85, huttunenheikinmaa05} to find the onsets of an SEP event in different energy channels. Cumulative Sum (CUSUM) methods are a set of similar statistical methods that are employed in many industries as quality control schemes \citep{page_1954_cusum}. CUSUM methods accumulate the difference between consecutive entries of time series data. The first entry of the CUSUM function is usually set to zero. In traditional CUSUM methods, the quality of an industrial product is monitored as a function of time, and this quality may be some numerical value assigned to the product. CUSUM methods are designed to give an early warning in the case this numerical value rapidly deviates from a default range of values, and as such they are widely used in SEP onset determinations \citep[e.g.,][]{Paassilta2018}.

The traditional CUSUM method operates under the assumption that the variable under monitoring is a random variable that is normally distributed. Hence, mean and standard deviation of the distribution are the two critical parameters that dictate if and when CUSUM gives a warning of the process getting out of control. We consider the intensity time series measurements as analogous to the quality of an industrially manufactured product, and as such find the onset of an SEP event at the moment of time when intensity measurements start to noticeably deviate from the pre-event background intensity.
In the classic approach of the CUSUM method, the monitored variable is assumed to follow a normal distribution. Because measurements of SEPs are expected to be Poisson-distributed, we use the modified Poisson-CUSUM method. To find the onset of an SEP event, it is important to first accurately define the pre-event background of the observations by its mean intensity $\mu$ and standard deviation $\sigma$. These parameters define the control parameter $k$ that restricts the growth of the CUSUM function:
\begin{align}
    \mu_{d} &= \mu + n \sigma \\
    k &= \frac{\mu_{d} - \mu}{\sigma(\ln(\mu_{d}) - \ln(\mu))},
\end{align}
\noindent where $\mu_{d}$ is an uncertainty limit for the background variations and $n$ is a natural number that is usually set to $n=2$. Finally, the dimensionless CUSUM function is calculated with normalized intensity measurements as follows:
\begin{align}
    I_{n} &= \frac{I - \mu}{\sigma} \\
    C[0] &= 0 \\
    C[i] &= \max(0, I_{n} - k + C[i-1]),
\end{align}
where $I$ is the measured intensity, $I_{n}$ is the normalized deviation from the mean intensity of the pre-event background, and $C$ is the CUSUM function. One can see from the definition of $C$ that it accumulates the difference of $I_{n}$ and $k$ as long as $I_{n}$ stays larger than $k$. When the values of $C$ exceed a defined hastiness threshold $h$, the method starts counting consecutive entries of $C$ until a predefined value of consecutive $C>h$ entries is reached. At that point, the method backtracks the same amount of data points, and marks the found data point as the onset of an event. The amount of consecutive $C>h$ entries that the method counts is a freely changeable parameter called \texttt{cusum\_window}. It is important to note that this parameter corresponds to the number of data points, and not to the time that has passed since $C$ exceeded $h$. The default value of \texttt{cusum\_window} $=30$ is chosen so that it corresponds to 30 minutes of continuous threshold-exceeding intensity measurements for a typical 1-minute time resolution. Following \citet{huttunenheikinmaa05}, we define $h$ such that it is by default set to $h=1$, but if $k$ is large, which in the context of Poisson-CUSUM means that $k \geq 1$, then it is set to $h=2$. Choosing $h$ to be sufficiently small is imperative for identifying the onset of the event as early as possible.

\begin{figure}[hp!]
\begin{center}
\includegraphics[height=0.91\textheight]{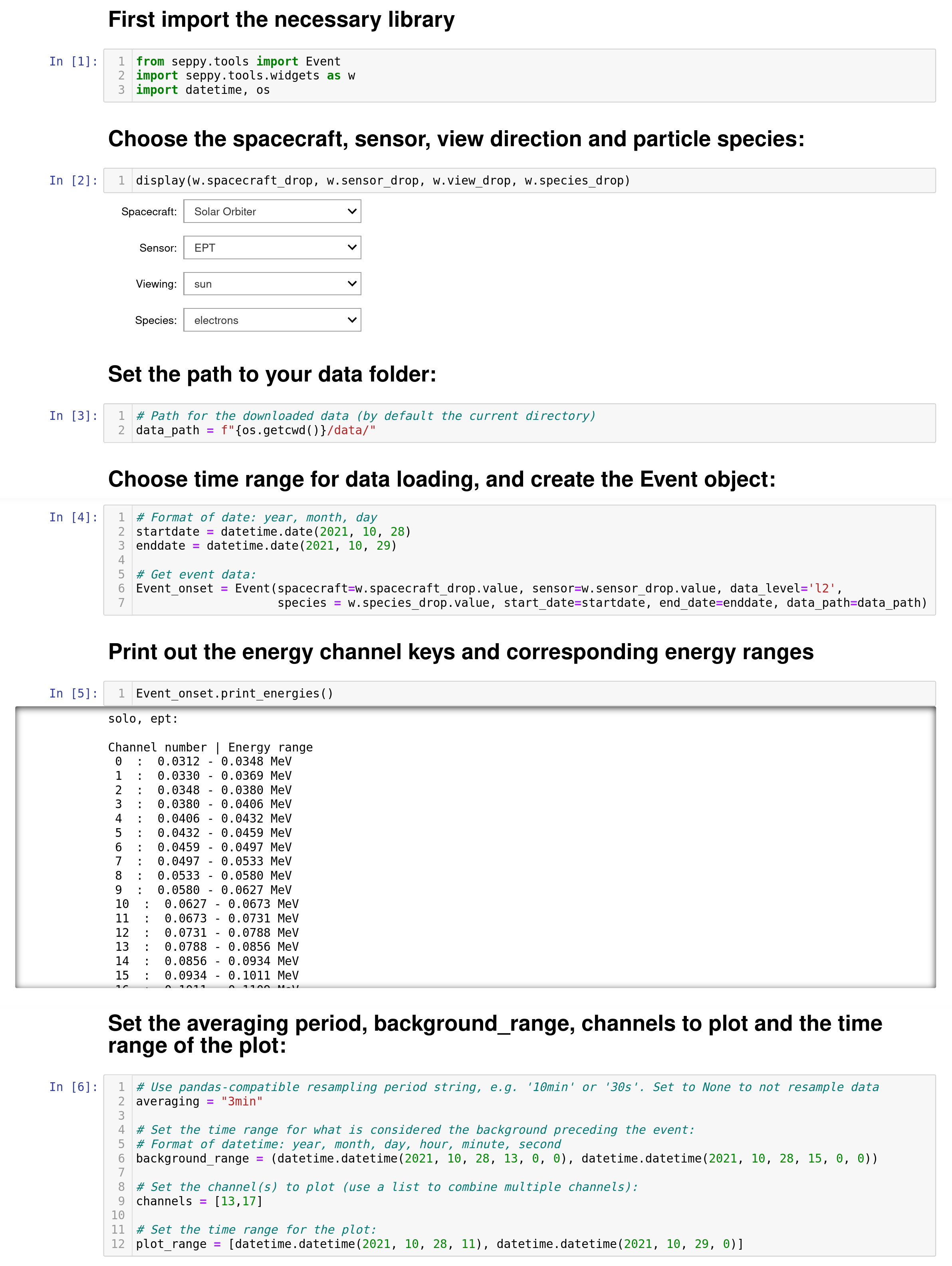}
\end{center}
\caption{The user interface of most Notebooks (here for the onset determination) consists of a simple drop-down menu accompanied by only a few lines of declaring variables. There are short instructions or a description between every cell that guide the user. This set of inputs yields an output shown in Fig.~\ref{fig:nb_onset_output}.} \label{fig:nb_onset_input}
\end{figure}

The onset time determination tool is built in Python, and the user interface is implemented in a Jupyter Notebook for easy usage. After importing the necessary libraries, the user needs to choose the observing spacecraft, measuring sensor, viewing direction (if the instrument provides multiple ones), and the particle species from drop-down menus (shown in Fig. \ref{fig:nb_onset_input}). After that, a data path can be defined to which data files should be saved to or loaded from. By default, this is set to the directory where the notebook is located. Finally, the user needs to specify the start and end of the time range that should be loaded. With these defined parameters, an \texttt{Event} object is initialized that contains the corresponding data and provides different functions that can be applied to it. Creating the object also automatically downloads the necessary data using the data loader functions presented in Sect.~\ref{sect:data_loaders} in case it is not found in the designated data directory.

\begin{figure}[ht!]
\begin{center}
\includegraphics[width=\textwidth]{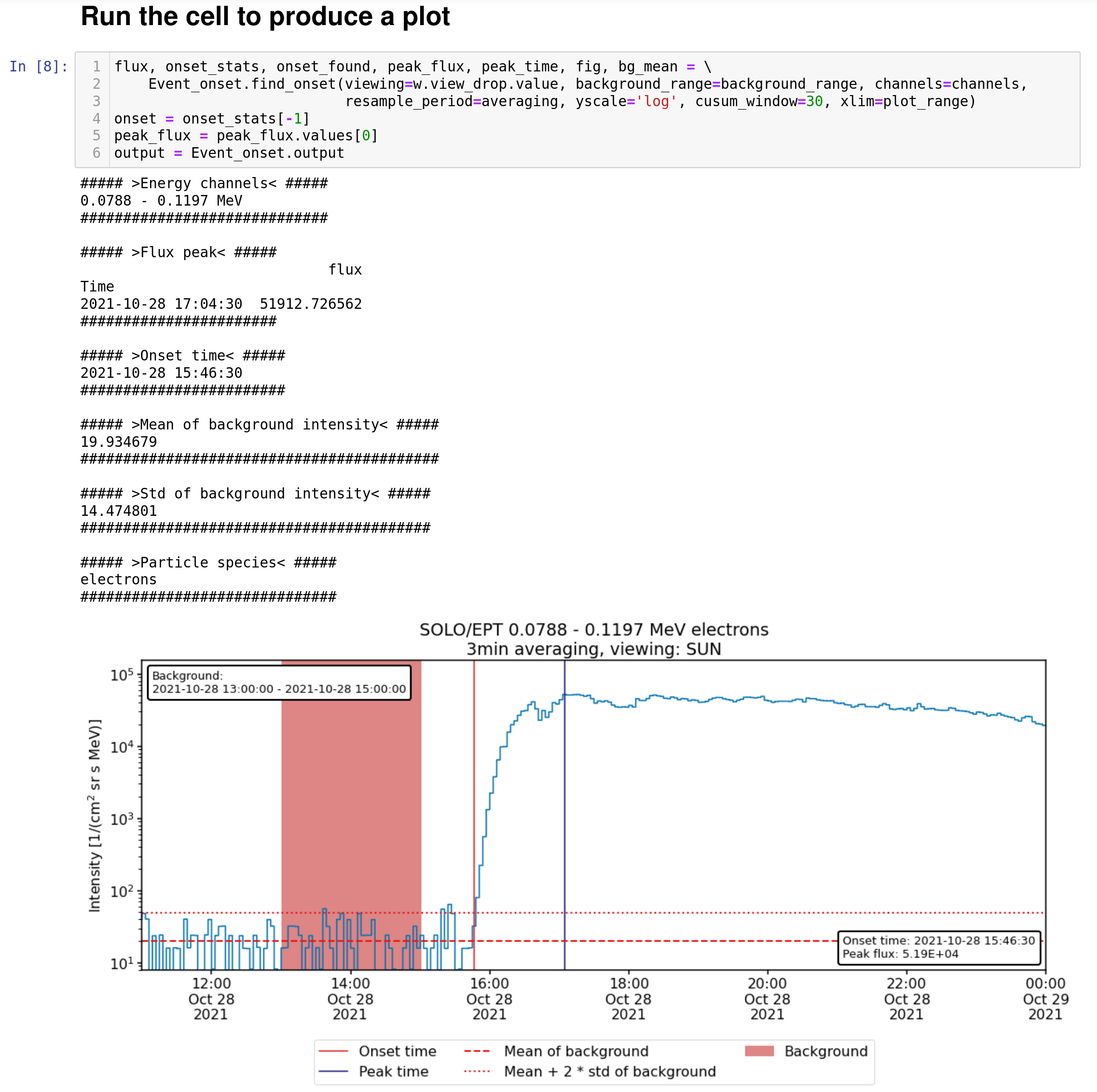}
\end{center}
\caption{Example output for the SEP onset analysis Notebook for the SEP event on 28 October 2021 as observed by Solar Orbiter/EPT, based on the inputs shown in Fig.~\ref{fig:nb_onset_input}. The red shaded area defines the pre-event background period, and the red and blue vertical lines indicate the found onset time and the peak intensity time, respectively. The intensity time series has been averaged to 1-minute resolution.} \label{fig:nb_onset_output}
\end{figure}

Once the object is created, the onset analysis can be run by defining four different parameters: \texttt{averaging}, \texttt{background\_range}, \texttt{channels}, and \texttt{plot\_range}. \texttt{averaging} is a parameter that defines the time-averaging of the intensity data. It accepts a Pandas-compatible time string, e.g., "1min" or "30s" for 1-minute or 30-second time resolution, respectively. It can also be set to \texttt{None}, in which case the intensity data is not time-averaged at all. It is noteworthy that data cannot be averaged to a finer resolution than its original measurement. \texttt{background\_range} defines the time window that is considered to be the pre-event background, and \texttt{plot\_range} defines the limits of the time-axis of the entire plot. The analysis can be run on a combination of multiple channels by giving the start and end indices of the channels as an input. This way, all channels in between are combined to a single energy channel, according to their respective energies and intensity measurements. It is of course also possible to run the onset analysis on only a single channel by just providing a single index as the input. The list of available energy channels for the chosen instrument can be obtained by running the function \texttt{Event.print\_energies()}. The output information and figure that are produced with the input parameters defined in Figure \ref{fig:nb_onset_input} are displayed in Figure \ref{fig:nb_onset_output}.

The CUSUM method is very effective in finding the time when the monitored variable suddenly changes drastically. While it still performs better than the 3-sigma method for gradual SEP rise phases, it can be challenging to find a correct onset time in such cases. Usually, a longer time averaging is recommendable then. Furthermore, a 
rational choice of the background interval is imperative for the method to work correctly, so the user needs to be careful in assessing whether to include transient structures that do not necessarily represent the actual background into \texttt{background\_range}.

\subsection{Dynamic spectrum}
\label{sect:spectra}

Whereas usual time series data is presented as a two-dimensional representation of intensity as a function of time in a single energy channel, a dynamic spectrum presents all the energy channels of an instrument in a single plot. This allows one to determine, e.g., confirmation of velocity dispersion in an event immediately with only visual inspection, something that would otherwise become apparent after finding the onset time in a number of energy channels.

The Dynamic Spectrum tool plots the dynamic spectrum of a single instrument for a given time range. The plot is initialized with a grid that has time bins on the x-axis and energy bins on the y-axis. The bins in the grid are then colored according to $I \cdot E^{2} $, where $E$ is the mean energy of the corresponding channel and $I$ is the observed intensity at that time and energy bin. Physically, $I\cdot E^2$ represents the energy flux per logarithmic energy band carried by the SEPs. The motivation behind plotting this quantity instead of simply $I$ is that this way the particle spectrum is flattened. We apply this procedure to account for the drastically varying intensity levels present in one energy spectrum, which can easily lead to poor visibility of the lowest intensities (observed at the highest energies) in a dynamic spectrum.

To plot the dynamic spectrum of an instrument, the user only needs to initialize the \texttt{Event} object presented in Sect.~\ref{sect:onsets}. There are only two optional parameters that the user can change: \texttt{averaging} works exactly as it does in onset analysis notebook, with a Pandas-compatible time string, and \texttt{cmap} accepts the name of one of Matplotlib's \citep{Hunter2007} \href{https://matplotlib.org/stable/tutorials/colors/colormaps.html}{colormaps} as a string. By default, \texttt{averaging} is set to \texttt{None} and \texttt{cmap} is set to \texttt{"magma"}.

\begin{figure}[ht!]
\begin{center}
\includegraphics[width=\textwidth]{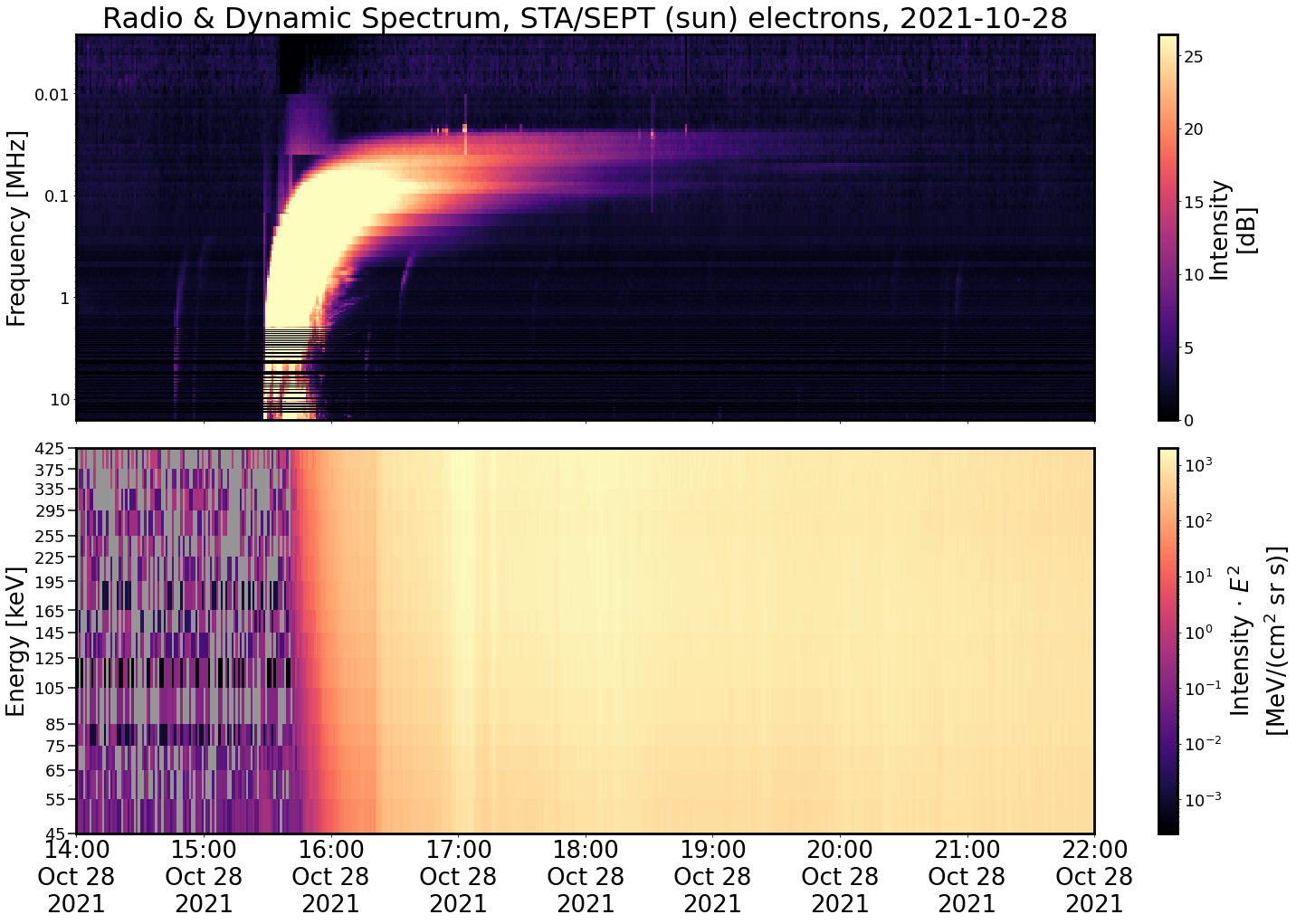}
\end{center}
\caption{The output of the dynamic spectrum tool for the 28 October 2021 SEP event with radio plotting enabled. The upper panel shows the radio spectrum as observed by STEREO-A/SWAVES and the lower panel the dynamic electron spectrum of STEREO-A/SEPT.}\label{fig:stereo_spectra}
\label{fig:spectra}
\end{figure}

Furthermore, the user also has the option to accompany the dynamic spectrum with a radio spectrum, as is shown in Fig. \ref{fig:spectra}. The radio spectrum in its current version only supports the SWAVES instruments onboard STEREO-A and B. It is planned to add support for the Wind/WAVES instrument in a future version. The maximum intensity in each spectrum is set to 70\% of the maximum intensity level of the time window specified. This is done to avoid saturation and to enhance weaker signals. The radio spectrum can be used to show the onset of radio bursts at low frequencies ($<10$~MHz), local Langmuir waves and electrostatic disturbances at the spacecraft.

\subsection{Visual Time Shift Analysis (TSA)}
\label{sect:tsa}
\begin{figure}[ht!]
    \centering
    \begin{minipage}{0.45\textwidth}
        \centering
        \includegraphics[width=1.0\textwidth]{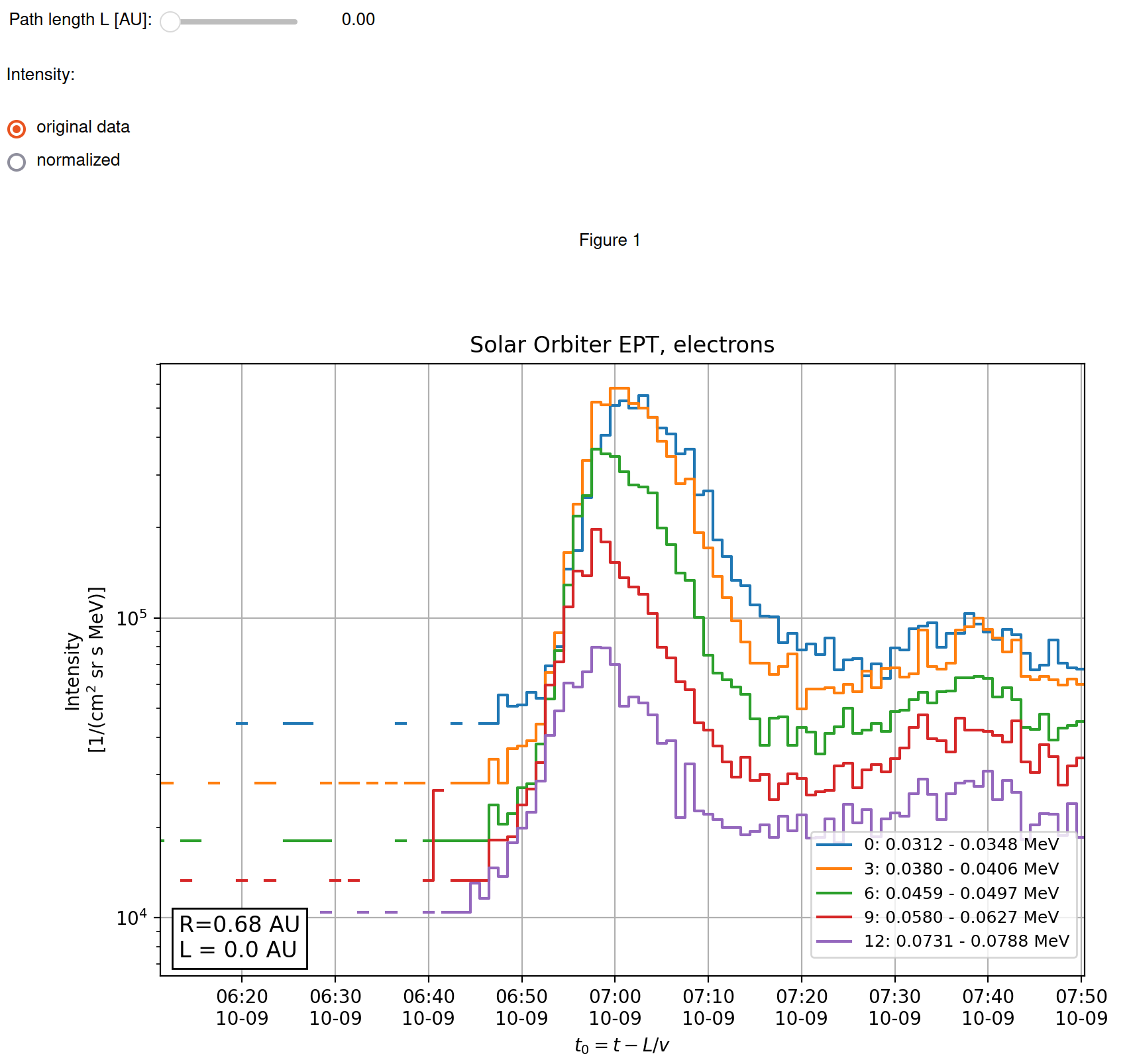}
    \end{minipage}\hfill
    \begin{minipage}{0.45\textwidth}
        \centering
        \includegraphics[width=1.0\textwidth]{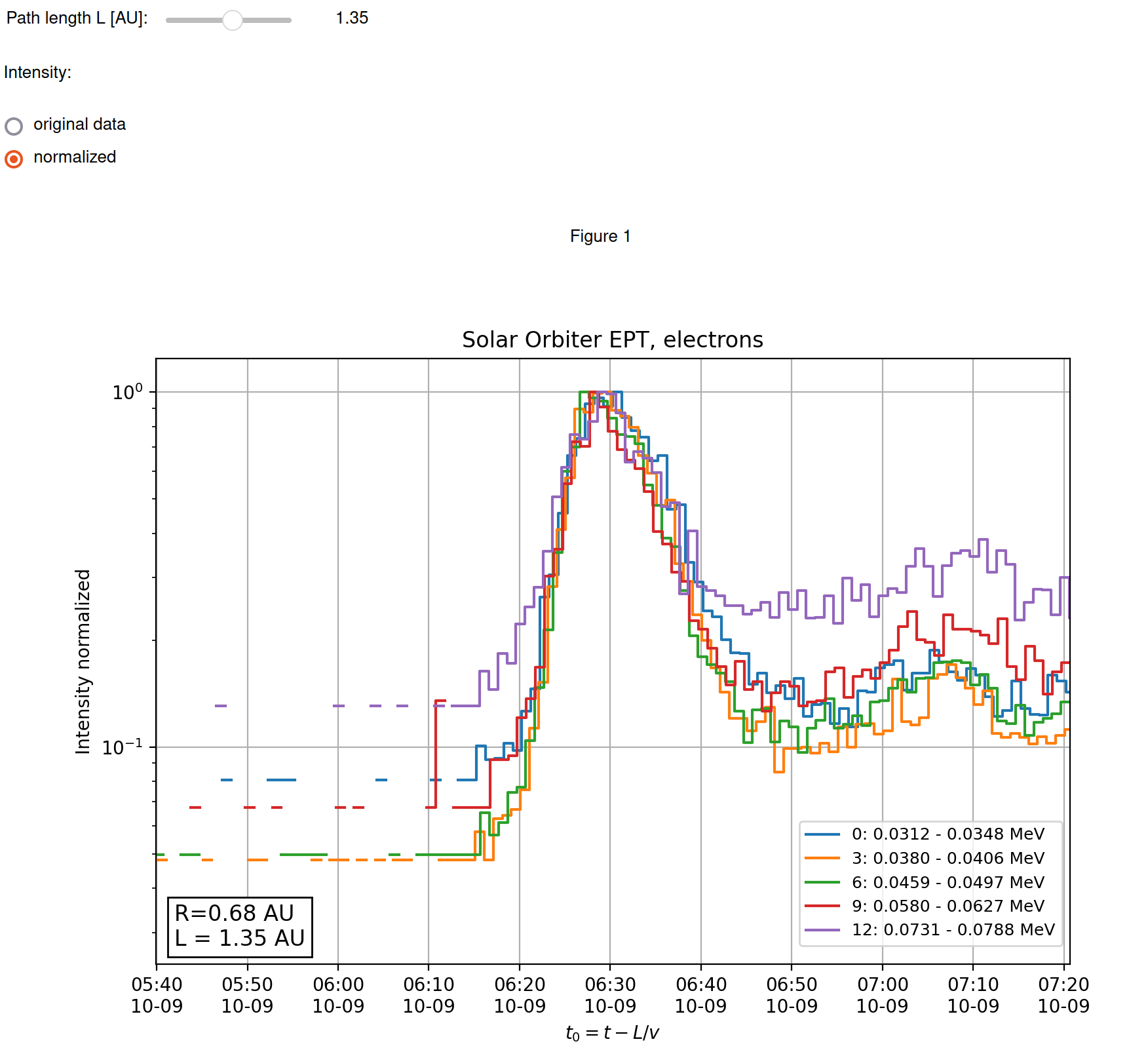}
    \end{minipage}
            \caption{Unaltered (left) and time-shifted (right) electron intensity time series of the 9 October 2021 SEP event observed by Solar Orbiter/EPT. The bottom left panels inside both plots show the heliocentric distance (R) of the spacecraft at the time of the event and the path length (L) that corresponds to the time-shift. The time-shifted plot on the right has been normalized to the maximum intensity of the event and time-shifted by eye so that the rising flanks of the different channels align, yielding a travelled path length of 1.35 AU. At the time of the event Solar Orbiter was located at a distance of 0.68 AU from the Sun, which corresponds to a path length of 0.73 AU, assuming a nominal Parker spiral with solar wind speed of 400 km/s}
            \label{fig:tsa}
\end{figure}

Velocity Dispersion Analysis (VDA) is a method to estimate the path length travelled by SEPs and the time when these SEPs were injected into the interplanetary medium. Often a VDA is not possible, for example, when the onset times in different energy channels do not show significant differences or when the onset times cannot be identified with high accuracy. For such cases, an alternative can be to use the rising phase at a fraction of the peak intensity instead of the classic onset of the event \citep[e.g.,][]{Zhao_2019}. VDA also does not take into account the pitch-angle scattering of particles in the interplanetary medium, which is why care has to be exercised in its application \citep{Laitinen_2015}. To still infer an SEP injection time, an alternative method, the Time Shift Analysis \citep[TSA, e.g.][]{vainio13_tsa} can be used, which usually employs only a single energy channel. Therefore, the information on the path length cannot be inferred. In our tool, we combine features of both methods by applying a TSA to a number of energy channels together.
The basic assumption behind this tool is that the particles propagated along a common path length. 
The interactive time shift analysis tool presented here shows the particle intensities of the different energy channels of the same instrument as time series in a single plot (see Fig. \ref{fig:tsa} left). The tool then allows to interactively apply different path lengths to the data. For each path length the time profiles of all chosen energy channels are shifted backwards in time corresponding to the particle's travel time along the chosen path length: 
\begin{equation}
    t_{0} = t - \frac{L}{v},
\end{equation}
where $t$ is the initial timestamp of an observation without a shift, $L$ is the assumed path length (i.e., along the interplanetary magnetic field line), and $v$ is the speed of the particles. By varying $L$, the time series data is shifted differently far backwards. The plot is updated automatically during runtime. Since different energy channels detect particles of different speeds, the plotted time series of different energy channels will be shifted at different rates. We note that all energy channels cover a finite energy range, so there will always be a range of speeds within the particle population in any given energy channel. Nonetheless, we take the geometric mean energy of a channel as an approximation to represent the kinetic energy of all particles in that energy range. The relativistic speed of a particle is then calculated according to:
\begin{equation}
    v = c \sqrt{1 - \bigg( \frac{mc^2}{E + mc^2} \bigg)^{2}},
\end{equation}
\noindent where $E$ is the mean energy of a channel, $m$ the mass of a particle, and $c$ is the speed of light. A reasonable path length, which explains the observed velocity dispersion, is found when the rise phases of all chosen energy channels collapse, i.e., when they have been shifted so that they align at the same time. This time corresponds to the inferred injection time. 
There is also the option to normalize all displayed energy channels to the maximum intensity in the considered time frame, which helps to align the rise phases in different energy channels. Figure \ref{fig:tsa} illustrates the usage of the tool by showing the unaltered intensity time series of multiple channels (left) next to normalized time series that have been time-shifted corresponding to a specific path length so that the onsets of the different channels align in a single onset (right).

\section{Discussion}
\label{sect:discussion}
In this paper, we present the first tools for analyzing SEP events that are provided through the SERPENTINE project to the community.
These Jupyter Notebooks are designed to alleviate the significant level of manual work that is an integral part of the study of any SEP event. They open up key pieces of energetic particle analysis for multi-spacecraft studies in an unprecedentedly easy way. The tools themselves require only a minimal amount of programming knowledge to utilize, while simultaneously giving the user access to powerful methods and visualization of data. There is, however, still room for some subjectivity in the interpretation of the results, e.g., for the onset and TSA tools. Also, the tools themselves require that the user is mindful of the methods and assumptions behind the curtain, in order to deliver scientifically valid analysis results. For example, it is still essential to read the corresponding data product descriptions and understand the specific caveats of every instrument before using the data. Figure~\ref{fig:data_loader_1} gives an example on the importance of understanding such instrument caveats. Thus, the tools are aimed at experts who are familiar with such kind of analyses.

The development of these tools is an ongoing endeavor, and the source code is open to everyone on GitHub. In the future, the user interface of the tools is planned to also support missions such as Wind and Parker Solar Probe, which so far are only partially supported. In addition, there are plans to include other data sets, such as X-ray and proton observations by GOES and radio data from other missions. Another goal is to provide access to higher level data products that are planned to be delivered by the SERPENTINE project, such as pitch-angle distributions or event catalogs. The existing tools will see further improvement, and new tools, such as VDA uncertainty estimation and SEP spectra analysis, are also under development. We encourage the scientific community to participate in the development by giving feedback about possible bugs and requested new functionalities via GitHub issues, but we are of course also open for direct communication via, e.g., email.

\section*{Conflict of Interest Statement}
%All financial, commercial or other relationships that might be perceived by the academic community as representing a potential conflict of interest must be disclosed. If no such relationship exists, authors will be asked to confirm the following statement: 

The authors declare that the research was conducted in the absence of any commercial or financial relationships that could be construed as a potential conflict of interest.

\section*{Author Contributions}
%The Author Contributions section is mandatory for all articles, including articles by sole authors. If an appropriate statement is not provided on submission, a standard one will be inserted during the production process. The Author Contributions statement must describe the contributions of individual authors referred to by their initials and, in doing so, all authors agree to be accountable for the content of the work. Please see  \href{https://www.frontiersin.org/about/policies-and-publication-ethics#AuthorshipAuthorResponsibilities}{here} for full authorship criteria.
CP, JG, ND, EA, and RV discussed the tools.
JG wrote the software for data loaders. CP, AY-L, EA, and JG wrote the software for onset determination. CP wrote the software for dynamic spectra and visual TSA. DEM wrote the software for radio data. 
AY-L, DP, and SV tested the software and provided critical comments for the development.
CP, JG, and ND prepared the first paper draft, and all authors were involved in the preparation of the final manuscript.
All authors revised the manuscript before submission.

\section*{Funding}
This study has received funding from the European Union’s Horizon 2020 research and innovation programme under grant agreement No.\ 101004159 (SERPENTINE).
CP, JG, RV, DP, and EA acknowledge the support of Academy of Finland (FORESAIL, grants 312357, 336809, and 336807).
ND acknowledges the support of Academy of Finland (SHOCKSEE, grant 346902). DEM acknowledges support from the Academy of Finland (RadioCME, grant 333859). DP and EA acknowledge the ERC under the European Union’s Horizon 2020 research and innovation programme (Grant agreement No. 724391, SolMAG). EA acknowledges the support of the Academy of Finland (TRAMSEP, grant 322455).

\section*{Acknowledgments}
The authors would like to thank everyone who helped to improve the tools described in this manuscript by providing feedback or contributing to the various open-source projects that it utilizes. The authors thank the two reviewers for their valuable feedback.
%This is a short text to acknowledge the contributions of specific colleagues, institutions, or agencies that aided the efforts of the authors.

% \section*{Supplemental Data}
%  \href{http://home.frontiersin.org/about/author-guidelines#SupplementaryMaterial}{Supplementary Material} should be uploaded separately on submission, if there are Supplementary Figures, please include the caption in the same file as the figure. LaTeX Supplementary Material templates can be found in the Frontiers LaTeX folder.

\section*{Data Availability Statement}
The source code used in this study can be found online:
\begin{itemize}
    \item SEPpy Python package code in GitHub repository     \href{https://github.com/serpentine-h2020/SEPpy}{SEPpy}, archived at \url{https://doi.org/10.5281/zenodo.7216077}
    \item solo-epd-loader Python package code in GitHub repository \href{https://github.com/jgieseler/solo-epd-loader}{solo-epd-loader}, archived at \url{https://doi.org/10.5281/zenodo.7100451}
    \item Python code and example Jupyter Notebooks in GitHub repository \href{https://github.com/serpentine-h2020/serpentine}{serpentine}, archived at \url{https://doi.org/10.5281/zenodo.7221070}
\end{itemize}
% % Please see the availability of data guidelines for more information, at https://www.frontiersin.org/about/author-guidelines#AvailabilityofData

% Define journal abbreviations for bibliography
\def\aj{AJ}%
          % Astronomical Journal
\def\actaa{Acta Astron.}%
          % Acta Astronomica
\def\araa{ARA\&A}%
          % Annual Review of Astron and Astrophys
\def\apj{ApJ}%
          % Astrophysical Journal
\def\apjl{ApJ}%
          % Astrophysical Journal, Letters
\def\apjs{ApJS}%
          % Astrophysical Journal, Supplement
\def\ao{Appl.~Opt.}%
          % Applied Optics
\def\apss{Ap\&SS}%
          % Astrophysics and Space Science
\def\aap{A\&A}%
          % Astronomy and Astrophysics
\def\aapr{A\&A~Rev.}%
          % Astronomy and Astrophysics Reviews
\def\aaps{A\&AS}%
          % Astronomy and Astrophysics, Supplement
\def\azh{AZh}%
          % Astronomicheskii Zhurnal
\def\baas{BAAS}%
          % Bulletin of the AAS
\def\bac{Bull. astr. Inst. Czechosl.}%
          % Bulletin of the Astronomical Institutes of Czechoslovakia 
\def\caa{Chinese Astron. Astrophys.}%
          % Chinese Astronomy and Astrophysics
\def\cjaa{Chinese J. Astron. Astrophys.}%
          % Chinese Journal of Astronomy and Astrophysics
\def\icarus{Icarus}%
          % Icarus
\def\jcap{J. Cosmology Astropart. Phys.}%
          % Journal of Cosmology and Astroparticle Physics
\def\jrasc{JRASC}%
          % Journal of the RAS of Canada
\def\mnras{MNRAS}%
          % Monthly Notices of the RAS
\def\memras{MmRAS}%
          % Memoirs of the RAS
\def\na{New A}%
          % New Astronomy
\def\nar{New A Rev.}%
          % New Astronomy Review
\def\pasa{PASA}%
          % Publications of the Astron. Soc. of Australia
\def\pra{Phys.~Rev.~A}%
          % Physical Review A: General Physics
\def\prb{Phys.~Rev.~B}%
          % Physical Review B: Solid State
\def\prc{Phys.~Rev.~C}%
          % Physical Review C
\def\prd{Phys.~Rev.~D}%
          % Physical Review D
\def\pre{Phys.~Rev.~E}%
          % Physical Review E
\def\prl{Phys.~Rev.~Lett.}%
          % Physical Review Letters
\def\pasp{PASP}%
          % Publications of the ASP
\def\pasj{PASJ}%
          % Publications of the ASJ
\def\qjras{QJRAS}%
          % Quarterly Journal of the RAS
\def\rmxaa{Rev. Mexicana Astron. Astrofis.}%
          % Revista Mexicana de Astronomia y Astrofisica
\def\skytel{S\&T}%
          % Sky and Telescope
\def\solphys{Sol.~Phys.}%
          % Solar Physics
\def\sovast{Soviet~Ast.}%
          % Soviet Astronomy
\def\ssr{Space~Sci.~Rev.}%
          % Space Science Reviews
\def\zap{ZAp}%
          % Zeitschrift fuer Astrophysik
\def\nat{Nature}%
          % Nature
\def\iaucirc{IAU~Circ.}%
          % IAU Cirulars
\def\aplett{Astrophys.~Lett.}%
          % Astrophysics Letters
\def\apspr{Astrophys.~Space~Phys.~Res.}%
          % Astrophysics Space Physics Research
\def\bain{Bull.~Astron.~Inst.~Netherlands}%
          % Bulletin Astronomical Institute of the Netherlands
\def\fcp{Fund.~Cosmic~Phys.}%
          % Fundamental Cosmic Physics
\def\gca{Geochim.~Cosmochim.~Acta}%
          % Geochimica Cosmochimica Acta
\def\grl{Geophys.~Res.~Lett.}%
          % Geophysics Research Letters
\def\jcp{J.~Chem.~Phys.}%
          % Journal of Chemical Physics
\def\jgr{J.~Geophys.~Res.}%
          % Journal of Geophysics Research
\def\jqsrt{J.~Quant.~Spec.~Radiat.~Transf.}%
          % Journal of Quantitiative Spectroscopy and Radiative Trasfer
\def\memsai{Mem.~Soc.~Astron.~Italiana}%
          % Mem. Societa Astronomica Italiana
\def\nphysa{Nucl.~Phys.~A}%
          % Nuclear Physics A
\def\physrep{Phys.~Rep.}%
          % Physics Reports
\def\physscr{Phys.~Scr}%
          % Physica Scripta
\def\planss{Planet.~Space~Sci.}%
          % Planetary Space Science
\def\procspie{Proc.~SPIE}%
          % Proceedings of the SPIE

\bibliographystyle{Frontiers-Harvard} %  Many Frontiers journals use the Harvard referencing system (Author-date), to find the style and resources for the journal you are submitting to: https://zendesk.frontiersin.org/hc/en-us/articles/360017860337-Frontiers-Reference-Styles-by-Journal. For Humanities and Social Sciences articles please include page numbers in the in-text citations 
\bibliography{references}

\end{document}